# Quantum Randi Challenge

Sascha Vongehr

National Laboratory of Solid State Microstructures and College of Engineering and Applied Sciences, Nanjing University, Jiangsu, P. R. China

Observed violations of Bell type inequalities exclude all relativistic micro causal ("local"), counterfactual definite ("real") hidden variable models of nature. The therefore inevitable further relativization of our concept of reality triggers a growing pseudoscientific resistance against quantum mechanics.
I define Didactic Randi Challenges (DRC) via five characteristics. These are challenges which, according to the laws of nature, are impossible to meet. They effectively refute pseudoscientific claims according to which the challenge could easily be met. DRC work by being known to exist while never having been overcome, despite the large rewards which would follow from meeting the challenge. Pseudoscience exploits well meaning engagement in argument to create the appearance of an expert dispute (sowing doubt). DRC decline to discuss "until the challenge is met", without solidifying the perception of establishment conspiracy. This requires transparency, thus DRC are efficient didactic tools.
The Quantum Randi Challenge (QRC) is a DRC designed to reject hidden variable models by simply teaching quantum mechanics; there is no bet or interaction with challengers. The QRC is a computer game that anybody can modify. The present version includes a simulation of true quantum behavior that violates Bell 99% of the time, hidden variables that violate the Bell and CHSH inequality with 50% probability, and ones which violate Bell 85% of the time when missing 13% anti-correlation. The DRC challenge is to modify the hidden variables so that the predicted quantum behavior arises, including anti-correlation. If such were possible, the presented programs would make it trivial to meet the challenge. This fact and the whole QRC can be taught to a wide audience via the presented heuristics. Demanding anti-correlation is argued to be superior to employing CHSH.

**6  Supplemental Material:   6 Figures showing Mathematica code and output**

# 1   Introduction: What is a Randi-Type Challenge?

The James Randi Educational Foundation (JREF) famously offers one million US dollars to anyone who can demonstrate paranormal abilities under laboratory conditions. Its existence has helped stem the spread of pseudoscience. I define a Didactic Randi Challenge (DRC) as one having the following necessary five characteristics:

**Randi-1**) It cannot be met (according to the established laws of nature).

**Randi-2**) If certain pseudoscientific claims were correct, it could be *easily* met.

**Randi-3**) Meeting the challenge would result in enormous rewards.

**Independence**) Judging whether the challenge is reasonable and whether it has been met is independent from anything that could be discredited as 'establishment conspiracy', for instance scientific peer-review or a single foundation.

**Didactic Transparency**) The necessity of Randi-2 and Independence demand transparency. Everything must be accessible to an educated lay audience to such a degree that the challenge is ideally left to that audience, for example judging whether the challenge has been met.

The original James Randi challenge fulfills only the first three criteria, but it already proved that challenges of this nature can be an effective tool. The five characteristics allow the following uses:

**U1**) Educators can point to the bare existence of the challenge to contest pseudoscience. The challenge having not been overcome in spite of items Randi-2 and Randi-3 argues that the claims of pseudoscience are wrong, convincing even many of those who cannot grasp the intricate details of the issue at hand. For instance, understanding that there can be a trivial error hidden behind the smoke screen of some highly complex calculation that claims to "disprove Bell" can be very difficult. However, the fact that the "anti-Bellist" does not go ahead and meet the corresponding challenge, which would bring her undying fame, with or without the approval of the scientific establishment, is a powerful argument that the anti-Bellist's theory cannot deliver what she claims.

  **U2**) The existence of the challenge allows scientists to refuse to enter into rhetoric arguments that mainly serve to provide pseudoscience a platform to promote itself. All communication is postponed until after the challenge is met. This aspect is important because one aim of pseudoscience, for example "intelligent design", is to spread doubt and construct the appearance of a controversy among experts, giving lay persons the impression that well-established science is in dispute. Well-meaning engagement in 'debates' backfires by supporting the deception. Often the mere exposure is desired and the debate perpetuated endlessly.

  **U3**) Didactic Transparency makes a DRC an efficient didactic tool that not only saves the time otherwise wasted refuting pseudoscience, but ideally is a fun activity that autonomously teaches the issue and the particular DRC self-contained and parallel to any curriculum.

  The general concept can be applied to counter pseudoscientific claims against quantum mechanics (QM) while simply teaching QM. This article describes the issues that the hereby officially announced Quantum Randi Challenge (QRC) addresses, and how it is ensured that the QRC has the crucial five DRC characteristics. This paper is still somewhat addressed at expert readers, because their support can promote the QRC to IT professionals and artists who can help perfecting it. However, this paper contains and for now *is* the QRC.

## 2   Pseudoscience against Quantum Physics

Pseudoscience often misrepresents QM to sell magic medical cures or argue precognition. Such is not our concern. We are concerned with the increasingly vocal *rejection* of QM especially by people who otherwise defend science. QM has been experimentally confirmed to astounding accuracy. Applications like quantum cryptography (Ekert 1991)[1] are based on superposition of states. Superposition is proven to be non-classical by the experiments and theory around the Einstein-Podolsky-Rosen (EPR) (Einstein 1935)[2] paradox and John Bell's famous inequality (Bell 1964)[3]. Uncertainty and quantization *could* emerge from classical substrates, but quantum superposition is not compatible with "local realism". "Local" stands for Einstein-locality (relativistic micro causality, sometimes distinguished from Einstein-separability). "Realism" is at times restricted to 'counterfactual definiteness'. This ill-defining of "realism", as well as 'spooky' non-locality already contesting naïve realism, both suggests the term *naïve* or *direct* realism. *Direct/naïve realism*[4] naively takes how things seem as if directly (without criticism) taken from the senses. In physics, it supposes that objects with all their properties are a certain definite way '*really out there*', which includes 'localism'.

The violation of Bell inequalities in experiments (Aspect 1981, 1982)[5,6] has disproved all directly real models, for example non-contextual, possibly stochastic, hidden variables. Such hidden variables cannot violate Bell's inequality (Bell 1966)[7], variations of which, like the Clauser-Horne-Shimony-Holt (CHSH) inequality (Clauser 1969)[8], have been strongly violated by diverse experiments, most impressively with the closing of the so called communication loophole by (Weihs 1998)[9]. Discussing an eavesdropper's exploitation of the still open 'detection loophole' is important for secure key distribution protocols (Barrett 2005; Acin 2006)[10,11]. However, such sophistication is ill advised when publicly defending QM against those who aim to save naïve realism by exploiting the detection loophole in ad hoc ways. Nature cunningly exploiting loopholes to deceive us about being classical would imply it wanting to do so rather than being a mere classical mechanism. Sophisticated refutation can validate nonsense as profound genius which the establishment allegedly cannot grasp and therefore suppresses. Any fundamental (rather than merely technical) detection loophole would be based on so called "further fact" uncertainties like holography (quantum gravity) or Bell's fifth position (Gill 2002)[12]. It would thus be even less classical than known QM.

**2.1 Confusing the Scientific Method with different Realisms**

Why is there anti-QM pseudoscience in the scientific community? There are different interpretations of QM. Some accept Everett relativity (Everett 1957)[13] and many-worlds (DeWitt 1973; Deutsch 1997)[14,15] or merely *modal* realism (Lewis 1986)[16]; some despise talk about parallel worlds, but all serious contenders know that *direct* realism is excluded, because the apparent EPR non-locality would have to be an instantaneous correlation inside a direct realism (not so in many world models). Although one cannot use it to transport matter or information with superluminal velocities, it would be a form of faster than light physics. This "*spooky* interaction at a distance" was already quite 'unreal' to A. Einstein and in fact, increasingly it is the "realism" in "local realism" which is called into doubt. Doubting any kind of "realism" meets resistance especially among researchers in more applied fields like engineering and chemistry and among science literate lay persons.  *Scientific realism* is defended precisely in order to reject pseudoscience, but different realisms are not properly distinguished.  Relaxing realism triggers a wide spectrum of concerns from questioning personal identity and responsible agency to fear of cultural relativism.  History tells similar about the adoption of Einstein relativity, but Everett relativity is more severe.  The reaction against modern physics will grow along with the acceptance of that QM demands to modify realism.  People who otherwise defend science are caught up in it.

# 3   The Quantum Randi Challenge

The following will first point out why the QRC is a DRC.  Afterward, the involved physics is explained simply in order to facilitate its further conversion into a high-school level exposition, which would need more space and visualizations.

## 3.1   The QRC is a Didactic Randi Challenge

The QRC challenges all claims of that QM predictions can be reproduced by directly real models.  The QRC fulfills all the criteria for a DRC:

**Randi-1**) QM predicts the violation of Bell inequalities, which has been experimentally observed.  Bell and others have proven that suchlike cannot possibly arise within a directly real model. The challenge cannot be overcome.

**Randi-2**) a) The challenge is to reproduce only and *nothing else but* the behavior of the simplest setup known to violate Bell's inequality maximally, which starts with parallel detectors (zero relative angle) and allows only one other angle for each of the two detectors. Only a small number (exactly 800) entangled photon pairs are to be modeled.

  b) Any directly real model's behavior can in principle be realized by classical computers. That classical computers can model the local systems is the essence of direct realism: everything depends on locally present data; variables have definite values at any time, even if they change randomly. (Some objections will be rejected later.)

  c) The computer program is already provided via several versions with different directly real example models being available here. All that a challenger would need to do is merely to modify the hidden variables and/or measurement prescriptions in order to reflect her specific model. Because it is a directly real model, modification is trivial: Given any directly real model, turning the particular hidden variables and measurement prescriptions into instructions for a programmer is trivial (doing it in such a way that the Bell inequality is violated is of course impossible).

  **Randi-3**) Instant fame is assured. In fact, since a three-computer setup itself constitutes a classical physical system, a Nobel Prize would be entirely deserved for whoever modifies the hidden variables and/or measurement prescriptions in the program so that the Bell inequalities are violated much more than 50% of the time, namely as QM predicts (about 99 times out of 100 runs while preserving anti-correlation).

  **Independence**) A Bell inequality violating program, published on the World Wide Web as a simple multi-player game, would become almost instantly famous, without any chance for established physicists ("the conspiring establishment") to prevent it. The QRC itself aims for internet virulence in niche-communities in order to achieve several aims at once, one being that challengers' modified programs will be automatically checked by many people that are not connected with academic science.

**Didactic Transparency**) Starting with this paper, the QRC explains *itself* in such a way as to best ensure that all steps are transparent and nothing is left as insider knowledge. Future downloadable programs shall be interactive tutorials that are also Alice and Bob role games; even the set up of any used web-page should be explained and as easy and inexpensive as possible.

That the QRC is effective in terms of the uses U1-3 is indicated by the successes that an initial deployment of the QRC (in yet less well presented versions) has had in on-line physics communities, for example:

**U1**) The mere existence of the QRC has been successfully employed (Vongehr 2011)[17] to discredit a particular classical model in the eyes of a lay audience that was up to then unconvinced by more "professional" refutations.

**U2**) Strictly scientific refutations (Gill 2003; Grangier 2007; Moldoveanu 2011)[18,19,20] of that model have shown to fuel a vicious cycle, triggering more claims to be refuted again. This has further popularized the involved claims, some of which subsequently even found funding sources and book deals. The QRC succeeded in terminating the artificially created debates on several popular web portals.

## 3.2   The EPR setup and Inequalities

The simplest EPR setup has a source of pairs of photons in its center. The photons are separated by sending them along the *x*-axis to Alice and Bob, who reside far away to the left and right, respectively. Alice has a polarizing beam splitter (a calcite crystal), which has two output channels. Alice's photon either exits channel "1", which leaves it horizontally polarized, or channel "0", which leads to vertical polarization (relative to the crystal's internal *z*-axis). The measurement is recorded as $A = 1$ or 0, respectively. This works just like with Polaroid sunglasses, which also split light into two polarized 'outputs'. The sunglasses absorb one 'output' channel while the crystal outputs both into slightly different directions. Bob uses a similar beam splitter, so that there are four possible measurement outcomes (*A*,*B*) for every photon pair: (0,0), (0,1), (1,0), or (1,1).

### *3.2.1 From Anti-correlation to Sine-dependence*

Every photon pair is prepared in 'singlet state entanglement', meaning that if the crystals are aligned in parallel, only the outcomes (0,1) and (1,0), for short U (for "Unequal"), will ever result. This is called anti-correlation. Although it is not easy to explain with linear polarizations, the underlying reasons can be heuristically motivated with well known classical symmetries like angular momentum conservation. For example, if the photon-pair is prepared with zero overall rotation and Alice's photon is observed to be circularly polarized, meaning its electrical field vector rotates a certain way (say clockwise), then Bob's photon must rotate the opposite way, because the total rotation is still zero. This anti-correlation reflects the consistency of the behavior of photons with classical optics, which the photons give rise to and that lay persons can understand. (This does not claim to derive QM from classical physics. Photons are quantum!) If the crystals are at an angle $\delta = (\beta - \alpha)$ relative to each other (rotated around the *x*-axis), the outcomes depend on $\delta$. But how do they depend on $\delta$? Anti-correlation implies that if $\delta = 0$ and Bob observes $B = 1$, Alice will get $A = 0$. Alice's un-measured photon *behaves as if* polarized orthogonally to Bob's measured one. It must be stressed that we should not think of Alice's photon being actually flipped to a certain polarization direction, triggered by Bob's measurement! Such is untrue, because the light speed limit forbids any information from Bob's measurement to arrive at Alice's place in time for her measurement. Nevertheless, anti-correlation at $\delta = 0$ implies that Alice's photon behaves *as if* polarized orthogonal to Bob's measurement outcome. However, $\delta$ may not even have been selected yet, say if Alice is further away from the photon source than Bob and if she sufficiently delays choosing her angle. Therefore, it is naturally expected that her photon behaves *as if* polarized orthogonally to Bob's photon at *all* angles $\delta$. This in turn makes the behavior of every single photon completely equal to that known from polarizing sunglasses! Polarizing filters split the light's electrical field vector into orthogonal components. It is simple geometry of projections (casting shadows) that the two orthogonal components are proportional to $\sin(\delta)$ and $\cos(\delta)$. Energy is proportional to the square of the field vectors, so energy is conserved: $\sin^2(\delta) + \cos^2(\delta) = 1$. Lay persons can even check this $\sin^2(\delta)$ dependence with polarizing sunglasses (and a photo diode and voltmeter). Energy is directly proportional to the number of photons, thus the photons in the light obey these factors as their probabilities for reaching the output channels, or else classical optics as we know it would not arise. The photon ends up with a probability proportional to $\sin^2(\delta)$ at one of the output channels of the crystal, as is usual for any polarized photon that meets a polarization filter. Therefore, the outcomes (0,0) and (1,1), for short E (for "Equal"), occur in the proportion $\sin^2(\delta)$. It is worthwhile to explain this, because this very $\sin^2(\delta)$ is precisely what violates the Bell inequality.

### 3.2.2 The Inequality predicted by Quantum Mechanics

Every experiment starts with the preparation of a pair of photons. When the photons are about half way on their paths to the crystals, Alice randomly rotates her crystal to let $\alpha = a\,(\pi/8)$ with $a$ either 0 or 3, i.e. $a \in \{0, 3\}$, selected at random by her throwing a coin (or observing another quantum measurement). Bob adjusts his crystal similarly to $\beta = b\,(\pi/8)$ with $b \in \{0, 2\}$. No other angles may be considered in order to ensure Randi-2. The magnitudes of $\delta = (b - a)\,(\pi/8)$ are multiples of 22.5°, with $d = |b - a|$, i.e. $d \in \{0, 1, 2, 3\}$. Hence, there are four equally likely cases: With $N_{\text{Total}} = 800$ photon pairs, the angles are about $N_d \approx 200$ times in each of the four configurations $d$. I avoid probabilities and consider only actual counts (never potential ones) expressed in small integers $N$. The outcomes of all runs are counted by the $4*2 = 8$ counters $N_d(X)$, where $X \in \{E, U\}$. Anti-correlation leads to $N_0(E) = 0$ and $N_0(U) \approx 200$. Generally, it holds that

$$N_d(E) \approx N_d \sin^2(\delta), \qquad N_d(U) \approx N_d \cos^2(\delta). \qquad (1)$$

Apart from $N_0(E) = 0$, only three of these are important: $N_1(U) \approx 200 * \cos^2(-\pi/8) \approx 170$ alone is expected to be by 40 occurrences *larger* than the sum of $N_2(E) \approx 200 * \sin^2(\pi/4) \approx 100$ and the third number $N_3(U) \approx 200 * \cos^2(-3\pi/8) \approx 30$. We expect

$$N_1(U) > N_2(E) + N_3(U).$$

A simulation of 800 photon pairs (Supplemental Material Fig. 6) which enforces the discussed sine-dependence, leads on average only nine times out of 1000 runs to the unlikely coincidence of $N_1(U)$ being smaller than the right hand sum. The program shows that quantum mechanics predicts the Bell inequality to be violated with a probability of around 99% ("almost always") already with only 800 pairs. Restricting to exactly 800 pairs keeps the numbers small enough to be in a lay person's comfort zone. This prevents the pseudoscientific practice of creating "smokescreens" with large numbers. $N_3(E) > N_1(E) + N_2(U)$ is expected similarly but not necessary for the QRC.

### 3.3 Hidden Variables and the Bell Inequality

Let us try to model the experiment described with help of hidden variables (HV). A pair of table tennis balls is prepared, say instructions are written on them, and then split. Before the balls arrive, Alice and Bob randomly select angles. Each ball results in a measurement 0 or 1 according to the angle it encounters and the HV (e.g. instructions) it carries. We do not assume anything about the complexity of the HV, which may be as complex as desired. Direct realism means here that each ball is a directly real object having all necessary information locally with it. Nothing needs to depend additionally on angles selected far away. This models the fact that photons travel at the speed of light. Nothing travels faster than light, so the photons must know any shared HV already when they are created and they must take this information with them on their way. Didactic transparency demands that Einstein-locality is not '*independence between statistical correlations*' but '*I cannot catch that ball anymore*'.

Assume the HV instructions somehow prescribe "If $a = 0$, then $A = 0$", short "$A_0 = 0$". The ball at Bob's place cannot know which angle Alice has just adjusted. She *might* have gotten $a = 0$, and if so, Bob's measurement cannot be also 0 if he also has $b = 0$. Thus, the HV, however complex they may be, must prescribe the complementary information "$B_0 = 1$". Furthermore, $A_3$ and $B_2$ must be somehow prescribed by the HV, otherwise the occurrences $N_d(A,B)$ cannot reproduce the $\sin(\delta)$ dependence. In summary, the HV may be an infinite table or a complex formula, but they must *at least effectively* contain the prescription of their *degrees of freedom* ($A_3$, $B_0$, $B_2$). [Anti-correlation fixes $A_0$ to equal 1 − $B_0$, so it is unnecessary.] According to these three degrees of freedom, each pair of balls falls into only one of $2^3 = 8$ different classes, which we may index by $i = 4A_3 + 2B_0 + B_2$, so that $i$ is the result of taking $A_3B_0B_2$ as a binary number. For example, $N^2$ counts occurrences of (0, 1, 0). The total number of pairs is $\sum_{i=0}^{7} N^i = 800$ again ($i$ is an index, not a power). Notice that ($A_3$, $B_0$, $B_2$) are the *degrees of freedom* of the HV as they *determine measurement outcomes* (not the HV, which can have any complexity). Alice's measurement of $A_0$ cannot change the value of $A_3$, because if $A_0$ is measured, $A_3$ is not measured, but $A_3$ is the value in case $a = 3$ *is* measured. We do not assume any type of counterfactual definiteness that is not even classically required, so HV may change or become random along the way. However, preparing HV, say (0, 0, 0), and then have them change with 30% probability to (0, 0, 1) on Bob's side in case $b = 2$, means to prepare the degrees of freedom of the HV three times out of ten as (0, 0, 1), not (0, 0, 0).

Every pair encounters one of the four possible configurations of angles, hence $N^i = N^i_0 + N^i_1 + N^i_2 + N^i_3$. All choices of angles occur about equally often and the HV cannot bias the choice (they have not arrived yet when the angles are chosen). Hence, all $N^i_d$ are expected to be roughly equal to $N^i/4$, which seems trivial enough but is a most important step, namely the very and only step where Einstein-locality comes in (closing the "communication loophole"; the HV have not arrived when the angles are selected):

$$N^i_d \approx N^i/4 \qquad (2)$$

All the cases counted by $N^4_d$, $N^0_0$, $N^0_2$, $N^1_0$, $N^5_0$, $N^5_3$, and $N^6_1$ imply measurement outcome $(A,B) = (1,0)$. Equivalently, $N^0_1$, $N^0_3$, $N^1_3$, $N^2_1$, $N^2_2$, and $N^6_2$ correspond to (0,0), while $N^1_2$, $N^5_1$, $N^5_2$, $N^6_3$, $N^7_1$, and $N^7_3$ to (1,1). Finally, $N^1_1$, $N^2_0$, $N^2_3$, $N^6_0$, $N^7_0$, $N^7_2$, and the four $N^3{}_d$ correspond to (0,1). This enumerates all the 32 possible $N^i{}_d$ exhaustively. Let us rearrange: All the cases $N^3{}_d$, $N^4{}_d$, $N^0{}_0$, $N^0{}_2$, $N^1{}_1$, $N^2{}_3$, $N^5{}_3$, $N^6{}_1$, and $N^7{}_2$ imply outcome (U), while $N^0{}_1$, $N^0{}_3$, $N^1{}_2$, $N^1{}_3$, $N^2{}_1$, $N^2{}_2$, $N^5{}_1$, $N^5{}_2$, $N^6{}_2$, $N^6{}_3$, $N^7{}_1$, $N^7{}_3$ correspond to (E). In Section 3.2, the following three counters were important: $N_1(U) = N^1{}_1 + N^3{}_1 + N^4{}_1 + N^6{}_1 \approx (N^1+N^3+N^4+N^6)/4$, $N_2(E) \approx (N^1+N^2+N^5+N^6)/4$, and $N_3(U) \approx (N^2+N^3+N^4+N^5)/4$. Bell's inequality is here the mathematically trivial statement that $N^1+N^3+N^4+N^6$ is by $2(N^2+N^5)$ *smaller* than $N^1+N^2+N^5+N^6$ and $N^2+N^3+N^4+N^5$ added together. In other words, it is expected that:

$$N_1(U) \leq N_2(E) + N_3(U) \qquad (3)$$

Even if the hidden variables are deliberately chosen in cunning ways, this inequality is expected because it derives from the randomness of the measurement angles leading to Eq.(2). Therefore, the quantum experiment described in Section 3.2, where $N_1(U)$ alone is *larger* than the right hand sum by 40, cannot be described by any directly real model.

### *3.3.1 Violating Inequalities*

Simply not preparing any $i = 2$ or $i = 5$ pairs sets $N^2$ and $N^5$ equal to zero and ensures that the equals sign in Eq.(3) is expected. The random fluctuations around the equality then violate the Bell (as well as the CHSH) inequality Eq.(3) in half of all runs on average (Sup. Mat. Fig. 4). Quantum mechanics violates the inequality 991 times out of 1000 (with 800 photon pairs). Hidden variables models that violate "often" and are presented as an advance toward a revolutionary discovery should be rejected by pointing out that a choice of hidden variables which violates Bell 50% of the time has been presented here already, and is thus uninteresting.

The measurement procedures can try to "cheat" in order to get more than 50% violation. For example, if the HV prescribe $i = 1$, then $B_2 = 1$ may increase $N^1_1$ or $N^1_2$. Alice can avoid the increase of $N^1_2$ by reporting $A_0 = 0$ (as if $i = 3$), but that increases $N_0(E)$ in case Bob reports $B_0 = 0$. Keeping anti-correlation requires Bob to collude with Alice: he must agree with her strategy in advance in order to report $B_0 = 1$ instead, however, that increases $N^3_3$ as often as $N^1_2$ is decreased; nothing is gained. Only by violating anti-correlation can they violate Bell more than 50% of the time. This is similar for every other combination: At $i = 6$, Alice can avoid the increase of $N^6_2$ by misreporting $A_0 = 1$, but that increases $N_0(E)$ in case $b = 0$. Alice misreporting $A_0 = 0$ in case $i = 1$ (and $a = 0$ obviously) makes the model violate the Bell inequality about 85% of the time (CHSH 50%), however anti-correlation at equal angles is already only 87% on average (Sup. Mat. Fig. 5). Any cheating that wants to conserve anti-correlation must communicate the angle settings between the players (i.e. violate Einstein locality) and would be easily spotted in the computer realizations.

### 3.4 Computer Realization

Basic computer realizations of the discussed EPR setup with hidden variables are simple. Implemented in Mathematica™, most is of the code performs the statistical analysis (Sup. Mat. Fig. 1). The core algorithm with Bell's random hidden variables, here conveniently the degrees of freedom ($A_3$, $B_0$, $B_2$), consists of only five vital lines of code (Sup. Mat. Fig. 2). Running it, including statistical analysis, are all accomplished in under one second on a ten year old PC. Constructing hidden variables is for example accomplished by the line:

n = 800; Table[H[j, k] = If[Random[] < 0.5, 0, 1], {j, n}, {k, 3}]

A typical output is:

*"Anti-correlation at equal angles OK.*
*{113, 106, 94}*
*The Bell inequality predicts that the first number is smaller than the sum of the second and third numbers. It holds in all directly real models (local realism). On principle, all directly real models can be realized by modifying this computer realization. QM violates the Bell inequality 99 times out of 100 runs (assuming 800 photon pairs per run), which excludes directly real models."*

The task for a challenger, who claims she has a hidden variable model that can give rise to quantum behavior, is to merely modify the program according to her model. Any directly real model whatever needs modification of only three lines of the code, namely the construction of the hidden variables and the parts where the measurement is accomplished in Alice's and Bob's places. The rest, like the random choice of angles, must stay unaltered. For example, if a challenger believes in hidden photon polarizations, a random angle $\rho \in [0, 2\pi]$ may replace the previous hidden variables:

Table[H[j] = 2π Random[], {j, 800}]

Mathematica represents these angles to six digits behind the decimal point, which allows a finer resolution than any EPR experiment has achieved. Note that if the model assumes photons to live in a hidden, multidimensional space, several angles may be entered. The hidden variables may reflect for example the topological double covering of the SU(2) group by angles periodic in $4\pi$ instead of $2\pi$. Randomness may be abandoned in favor of a table describing 800 fixed objects.

The QRC has much didactic potential through this 'gaming'. If, for example, measurement outcomes are mistakenly believed to be only due to the probabilities as they are known from single photons at polarized filters (*classical* indeterminism), a random selection with probability $\cos^2(\alpha - \rho)$ may modify the measurement part of the program. An example for a thus modified program (Sup. Mat. Fig. 3) leads to the output:

*"75.3% Anti-correlation only. Model fails to describe anti-correlation when Alice and Bob happen to measure with the same angle. …"*

The simplicity of the program derives from the simplicity of the experimental setup, i.e. the restricted choice of angles, *not* from enforcing simplicity of the hidden variables, which may have any complexity. Anybody who has come up with a novel model would be able to modify the program according to it – a much easier task than thinking up models with exotic statistical correlations. If the modified program then indeed violates the Bell inequality, it would attract a great deal of attention. People would help turn it into an online multiplayer game as described in the remarks inside the program (ideally, with your help, the QRC will be soon a multiplayer game already, so it needs only few modifications). Already available entertainment multiplayer games are much more complex than the one envisioned here and a whole industry exists to program them and educate programmers. Such a game, distributed over three different computers (host server as photon source, Alice, and Bob) could cut internet communication at appropriate times to prevent cheating via artificial non-locality. It would then constitute a classical physical system, which if it violated Bell's inequality, would become known worldwide in a matter of weeks (ensuring Randi-3 and Independence).

### 3.5  Discussion

Games have been discussed before (Vaidman 2001, and refs therein)[21], even in form of computer simulations constituting a challenge (Gill 2003)[18] against dubious claims. The QRC is different.

### *3.5.1  Simulation, Computer Model, or Physical System*

Firstly, the QRC is not a simulation but would, in case hidden variables could violate the Bell inequality, be a classical physical system that does so. The described multi-player game computer setup constitutes a classical physical system; computers are physical! The equivalence of empirical classical physics and classical computation is guarantied by that all experimental observations have finite resolution due to experimental accuracy. The finite capacity of computer memory does therefore not present an obstacle if physics involves true continuums. Some have suggested that the hidden variables are "topological" and related to hyper spheres. This is irrelevant, because there is no difference to a computer about whether it calculates relations applicable to three dimensional Euclidian space or something else. Many geometries and topologies (e.g. black holes and worm holes and the SU(2) double covering that Fermions are susceptible to) have been modeled. Computers do not know which of those geometries they happen to compute in.

### *3.5.2  Bets, Statistical Thresholds, Winning Challengers*

The QRC is not a bet. It *refuses* interaction with challengers. It is irrational to try convince irrational challengers rationally. The QRC is explicitly about *refusing* interaction (U2) with people who insist on an agenda designed to discredit quantum mechanics. The QRC is not about whether the detection loophole may hide spooky superdeterminism. The QRC is addressed at those who can accept quantum mechanics as that particular theory that has been validated with unprecedented accuracy in for example high energy particle physics, optics, and quantum chemistry. That theory predicts anti-correlation for singlet states, and not that photons conspire to escape detection in just the right way to fool humans. There is no argument that empirical science can present against positions such as planted fossil records and it is not its task to attempt such.

The QRC rejects all statistical thresholds, where a bet is lost by sheer luck, because as far as quantum mechanics is known today (unitary, no gravity corrections), it allows challengers to win no matter how small the probability. Using only 800 photon pairs ensures that most models violate Bell's inequality sometimes if players try often enough. This teaches the randomness involved. "*Oh they never win against those odds*" is especially wrong in our quantum world, because they *do* win against any simple threshold in plenty of "parallel worlds". Didactic Transparency demands to admit this.

### 3.5.3  *Anti-correlation versus CHSH, Convexity*

This section is unnecessary for the wider audience, because it addresses the issue of why we refuse the CHSH. The CHSH is superior when discussing the detection loophole. It avoids the $\delta = 0$ angle and works for quite mixed, not maximally entangled states that have no anti-correlation at any of the set angles. This makes CHSH a bad choice for the QRC. Starting with the simple $\delta = 0$ setup and anti-correlation connects to well known optics and classical common cause correlation. After all, pseudoscience claims that quantum correlations are merely a 'more complicated form' of classical correlations, but not fundamentally more profound than Alice having the right sock of a pair of socks if Bob has the left sock. However, "quantum phenomena are more disciplined" (Peres 1978)[22] than even perfect classical common cause correlation can provide. The $\delta = 0$ situation with anti-correlation shows that the classical correlation is indeed present but obviously not the full issue. Sine-dependencies further violate the inequalities, and anti-correlation is merely sin(0) = 0. It facilitates insight and Didactic Transparency if such is not hidden via the CHSH. Avoiding the CHSH supports the view that the detection loophole is a technicality that can be narrowed further by improving detectors, while only the communication loophole is of crucial importance, because Einstein-locality is crucial.

  Some insist on the CHSH, because 95% reliable photon-pair emission, transmission, and coincidence-detection, which is sufficient for the CHSH to close the detection loophole, are conceivable, but perfect anti-correlation can never be ensured, neither by the photon-pair preparation nor by the angle settings. However, since it is agreed that the QRC is fine in case of perfectly precise angles for example, it is not reasonable to argue that it suddenly fails if we misalign an angle by just 0.01 degrees, thus destroying perfect anti-correlation. The quantum violation of the inequalities is much too large for this to be an issue and the *amount* of violation depends on the sin($\delta$), which is merely less obvious with the CHSH. Hidden variables can violate CHSH 50% of the time (Section 3.3, Sup. Mat. Fig. 4 and 5). Quantum mechanics violates the uncertainties 99% of the time (with 800 pairs) while keeping anti-correlation. The QRC is about accepting quantum mechanics much like we accept special relativity today, and demanding reproduction of what that theory predicts for the singlet states, which belong to the setup just as much as the specific angle settings. Experiments may perhaps never prove more than 70 Bell violations out of 100 trials, because of either technicalities or perhaps certain 'further facts' like 'Bell's fifth position' related to further complementarity principles (holographic black hole complementarity or Diosi-Penrose criterion) or the overall spin of the EPR setup limiting angle resolution, which are all 'even more quantum' rather than a retreat to classicality.

Bell's argument fails without the determinism due to demanding anti-correlation, and so any lenience about anti-correlation must argue convexity and fair sampling. By convexity, every indeterministic hidden variables theory can be replaced by a deterministic one without affecting the observed statistics (this has been implicitly argued above via the distinction between degrees of freedom and hidden variables). The CHSH allows lenience, but the QRC would lose Didactic Transparency. With the QRC, convexity is trivial: Anti-correlation at $\delta = 0$ is the full classical correlation, and any further randomness via "genuinely stochastic" hidden variables can at most lead to less correlation, not to yet "*more disciplined*" correlation. Anyway, variables with random functions being evaluated at the measurement locations are allowed in the QRC and have already been described in Section 3.4; they fail anti-correlation.

# 4   Acknowledgements

I thank Richard D. Gill (Mathematics Dept. University of Leiden, Netherlands) and Matthew S. Leifer (Physics Dept. University College London, GB) for discussing patiently and suggesting many important changes.

## 6 Supplemental Material

**Figure S1**: The Statistical Analysis module

```
(*This part cannot be modified.*)
AC := {{N₀ = Σⱼ₌₁ⁿ If[α[j] == β[j], 1, 0], N_E0 = Σⱼ₌₁ⁿ If[And[α[j] == β[j], A[j] == B[j]], 1, 0]};
  If[N_E0 > 0, N[100 - 100 * N_E0 / N₀] "% Anti-correlation only. Model fails to describe anti-correlation when Alice and Bob happen to measure with the same angle.",
    "Anti-correlation at equal angles OK."]}

BellT := {{N_U1 = Σⱼ₌₁ⁿ If[And[β[j] - α[j] == -π/8, A[j] ≠ B[j]], 1, 0], N_E2 = Σⱼ₌₁ⁿ If[And[β[j] - α[j] == π/4, A[j] == B[j]], 1, 0], N_U3 = Σⱼ₌₁ⁿ If[And[β[j] - α[j] == -3π/8, A[j] ≠ B[j]], 1, 0]},
  "The Bell inequality predicts that the first number is smaller than the sum of the second and third numbers.
    It holds in all directly real models (local realism).
    On principle, all directly real models can be realized by modifying this computer realization.",
  If[N_U3 + N_E2 < N_U1, "Bell's inequality is violated! Please play again.
    QM violates Bell's inequality roughly 99 times out of 100 (assuming 800 photon pairs per trial).",
    "QM violates the Bell inequality 99 times out of 100 runs (assuming 800 photon pairs per run), which excludes directly real models."]}

CHSH := {{N₃ = Σⱼ₌₁ⁿ If[β[j] - α[j] == -3π/8, 1, 0], N_E3 = Σⱼ₌₁ⁿ If[And[β[j] - α[j] == -3π/8, A[j] == B[j]], 1, 0],
  N₁ = Σⱼ₌₁ⁿ If[β[j] - α[j] == -π/8, 1, 0], N_E1 = Σⱼ₌₁ⁿ If[And[β[j] - α[j] == -π/8, A[j] == B[j]], 1, 0], N₂ = Σⱼ₌₁ⁿ If[β[j] - α[j] == π/4, 1, 0],
  E₀ = 2 N_E0/N₀ - 1, E₁ = 2 N_E1/N₁ - 1, E₂ = 2 N_E2/N₂ - 1, E₃ = 2 N_E3/N₃ - 1}; CHV = N[Max[Abs[E₀ + E₁ + E₂ - E₃], Abs[E₀ + E₁ - E₂ + E₃], Abs[E₀ - E₁ + E₂ + E₃], Abs[E₁ + E₂ + E₃ - E₀]]],
  If[CHV > 2, "CHSH inequality is violated!", "CHSH inequality is not violated."]}
```

**Figure S2**: The Bell hidden variables model with remarks on how it should be turned into a multi-player game in order to be a physical realization rather than simulation. Below the green rectangle is the output of a typical run.

```
(*These hidden variables may be modified at will but must be entirely computed on the host computer.*)
n = 800; Table[H[j, k] = If[Random[] < 0.5, 0, 1], {j, n}, {k, 3}];

(*This part must be computed on Alice's computer. Alice's angles α[j] must be computed before the hidden variables arrive and may not change
   afterward. The next formula may not be modified.*)
Table[α[j] = If[Random[] < 0.5, 0, 3] (π/8), {j, n}];
(*The next formula may be modified, but Alice's measurements A[j] can only be a table of 800 values ∈ {0,1}.*)
Table[A[j] = If[α[j] == 0, 1 - H[j, 2], H[j, 1]], {j, n}];

(*This part must be computed on Bob's computer. Bob's angles β[j] must be computed before the hidden variables arrive and may not change
   afterward. The next formula may not be modified.*)
Table[β[j] = If[Random[] < 0.5, 0, 2] (π/8), {j, n}];
(*The next formula may be modified, but Bob's measurements B[j] can only be a table of 800 values ∈ {0,1}.*)
Table[B[j] = H[j, If[β[j] == 0, 2, 3]], {j, n}];

(*This part cannot be modified. It can be computed on any computer.*)
AC
MatrixForm[BellT]
Clear[H, A, B, α, β]
```

{Anti-correlation at equal angles OK.}

{104, 70, 100}
The Bell inequality predicts that the first number is smaller than the sum of the second and third numbers.
  It holds in all directly real models (local realism).
  On principle, all directly real models can be realized by modifying this computer realization.
QM violates the Bell inequality 99 times out of 100 runs (assuming 800 photon pairs per run), which excludes directly real models.

**Figure S3**: An example for how a challenger who believes the photons to carry fixed polarization vectors **r** may modify the given, simplest example above. The analysis rejects it for its poor anti-correlation. The modification is almost trivial, although the model now includes vectors in coordinate space as hidden variables.

```
n = 800; Table[H[j] = 2 π Random[], {j, n}];

Table[α[j] = If[Random[] < 0.5, 0, 3] (π/8), {j, n}];
Table[A[j] = If[Random[] < (Cos[α[j] - H[j]])^2, 1, 0], {j, n}];

Table[β[j] = If[Random[] < 0.5, 0, 2] (π/8), {j, n}];
Table[B[j] = If[Random[] < (Cos[β[j] - (H[j] + π/2)])^2, 1, 0], {j, n}];

AC
MatrixForm[BellT]
Clear[H, A, B, α, β]
```

{78.2383 % Anti-correlation only. Model fails to describe anti-correlation when Alice and Bob happen to measure with the same angle.}

{145, 93, 84}

The Bell inequality predicts that the first number is smaller than the sum of the second and third numbers.
It holds in all directly real models (local realism).
On principle, all directly real models can be realized by modifying this computer realization.
QM violates the Bell inequality 99 times out of 100 runs (assuming 800 photon pairs per run), which excludes directly real models.

**Figure S4**: These hidden variables violate the Bell and CHSH inequalities in 50% of all runs.

```
n = 800; Table[H[j, k] = If[Random[] < 0.5, 0, 1], {j, n}, {k, 3}];
(*The next formulas get rid of N^2 and N^5 and thus saturate the Bell inequality to make it an equality at high n. Bell is thus violated 50 % of the time.*)
Table[i[j] = 4 H[j, 1] + 2 H[j, 2] + H[j, 3], {j, n}];
Table[H[j, k] = If[Or[i[j] == 2, i[j] == 5], 1 - H[j, 1], H[j, 1]], {j, n}, {k, 1}];

Table[α[j] = If[Random[] < 0.5, 0, 3] (π/8), {j, n}];
Table[A[j] = If[α[j] == 0, 1 - H[j, 2], H[j, 1]], {j, n}];

Table[β[j] = If[Random[] < 0.5, 0, 2] (π/8), {j, n}];
Table[B[j] = H[j, If[β[j] == 0, 2, 3]], {j, n}];

AC
MatrixForm[BellT]
CHSH
Clear[i, H, A, B, α, β]
```

{Anti-correlation at equal angles OK.}

{150, 110, 41}

The Bell inequality predicts that the first number is smaller than the sum of the second and third numbers.
It holds in all directly real models (local realism).
On principle, all directly real models can be realized by modifying this computer realization.
QM violates the Bell inequality 99 times out of 100 runs (assuming 800 photon pairs per run), which excludes directly real models.

{2.00995, CHSH inequality is violated!}

**Figure S5**: These hidden variables violate the Bell inequality 85% of the time (CHSH 50%) because they miss anti-correlation about 13% of the time.

```
n = 800; Table[H[j, k] = If[Random[] < 0.5, 0, 1], {j, n}, {k, 3}];
(*The next formulas get rid of N^2 and N^5 and thus saturate the Bell inequality to make it an equality at high n.*)
Table[i[j] = 4 H[j, 1] + 2 H[j, 2] + H[j, 3], {j, n}];
Table[H[j, k] = If[Or[i[j] == 2, i[j] == 5], 1 - H[j, 1], H[j, 1]], {j, n}, {k, 1}];

Table[α[j] = If[Random[] < 0.5, 0, 3] (π/8), {j, n}];
(*The next formula's red modification gets rid of N^1_2 and thus violates anti-correlation.*)
Table[A[j] = If[α[j] == 0, If[i[j] == 1, H[j, 2], 1 - H[j, 2]], H[j, 1]], {j, n}];

Table[β[j] = If[Random[] < 0.5, 0, 2] (π/8), {j, n}];
Table[B[j] = H[j, If[β[j] == 0, 2, 3]], {j, n}];

AC
MatrixForm[BellT]
CHSH
Clear[i, H, A, B, α, β]
```

{83.9806 % Anti-correlation only. Model fails to describe anti-correlation when Alice and Bob happen to measure with the same angle.}

{142, 76, 55}

The Bell inequality predicts that the first number is smaller than the sum of the second and third numbers.
  It holds in all directly real models (local realism).
  On principle, all directly real models can be realized by modifying this computer realization.
     Bell's inequality is violated! Please play again.
        QM violates Bell's inequality roughly 99 times out of 100 (assuming 800 photon pairs per trial).

{1.88235, CHSH inequality is not violated.}

**Figure S6**: The simulation of quantum behavior needs knowledge of the relative angle. This simulation has been run 1000 times. With 800 pairs, only 9 times is the Bell inequality not violated.

```
n = 800;
(*Random angles and random outcomes.*)
Table[α[j] = If[Random[] < 0.5, 0, 3] (π/8), {j, n}];
Table[A[j] = If[Random[] < 0.5, 0, 1], {j, n}];

(*Random angles also for Bob, but Bob's outcomes stay undetermined relative to Alice.*)
Table[β[j] = If[Random[] < 0.5, 0, 2] (π/8), {j, n}];

(*Bob's outcomes are correlated with Alice's angles.*)
Table[B[j] = If[Random[] < (Sin[β[j] - α[j]])^2, A[j], 1 - A[j]], {j, n}];

AC
MatrixForm[BellT]
CHSH
Clear[A, B, α, β]
```

{Anti-correlation at equal angles OK.}

{159, 95, 36}

The Bell inequality predicts that the first number is smaller than the sum of the second and third numbers.
  It holds in all directly real models (local realism).
  On principle, all directly real models can be realized by modifying this computer realization.
     Bell's inequality is violated! Please play again.
        QM violates Bell's inequality roughly 99 times out of 100 (assuming 800 photon pairs per trial).

{2.23468, CHSH inequality is violated!}

26